# Multichannel vectorial holographic display and encryption

Ruizhe Zhao[1], Basudeb Sain[2], Qunshuo Wei[1], Chengchun Tang[3,4], Xiaowei Li[5], Thomas Weiss[6], Lingling Huang[1,2\*], Yongtian Wang[1†], Thomas Zentgraf[2‡]

1. School of Optics and Photonics, Beijing Institute of Technology, Beijing, 100081, China
2. Department of Physics, University of Paderborn, Warburger Straße 100, 33098 Paderborn, Germany
3. Institute for Quantum Science and Engineering, Shenzhen, 518055, China
4. Department of Physics, Southern University of Science and Technology, Shenzhen, 518055, China
5. Laser Micro/Nano-Fabrication Laboratory, School of Mechanical Engineering, Beijing Institute of Technology, Beijing 100081, China
6. 4th Physics Institute and Research Center SCoPE, University of Stuttgart, Pfaffenwaldring 57, 70569 Stuttgart, Germany

**Abstract:** Since its invention, holography has emerged as a powerful tool to fully reconstruct the wavefronts of light including all the fundamental properties (amplitude, phase, polarization, wave vector and frequency). For exploring the full capability for information storage/display and enhancing the encryption security of metasurface holograms, smart multiplexing techniques together with suitable metasurface designs are highly demanded. Here, we integrate multiple polarization manipulation channels for various spatial phase profiles into a single birefringent vectorial hologram by completely avoiding unwanted cross-talk. Multiple independent target phase profiles with quantified phase relations that can process significantly different information in different polarization states are realized within a single metasurface. For our metasurface holograms, we demonstrate high fidelity, large efficiency, broadband operation, and a total of twelve polarization channels. Such multichannel polarization multiplexing can be used for dynamic vectorial holographic display and can provide triple protection for optical security. The concept is appealing for applications of arbitrary spin to angular momentum conversion and various phase modulation/beam shaping elements.

\* Email: huanglingling@bit.edu.cn
† Email: wyt@bit.edu.cn
‡ Email: thomas.zentgraf@upb.de

**Introduction:**

Holography based on metasurfaces has emerged as a promising candidate for applications in optical displays, optical storage and security by exhibiting unprecedented spatial resolution, enormous information capacity and a large field of view compared to traditional methods[1-11]. The practical realization of metasurface holograms relies on the encoding of holographic profiles on ultrathin nanostructures, which can possess strong light-matter interaction within an ultrashort distance[12-18]. Many efforts have been put into incorporating novel holographic-related functionalities such as beam shaping and ghost imaging into metasurfaces[19-22]. By utilizing the arbitrary design freedom of metasurfaces to simultaneously engineer the local wavelength, amplitude, phase, and polarization response, it might be possible to provide a flexible and compact platform to realize all types of vectorial holograms (with high-dimensional information) rather than scalar intensity information, which can surpass the limitations of natural materials such as liquid crystals or optical photoresists[23-26].

For the purpose of optimizing the tremendous information capability of metasurface holograms, multiplexing techniques are highly desired including color holograms[27-29], polarization multiplexing[30-34] and hybrid multiplexing[35,36]. Among the mentioned multiplexing techniques, polarization multiplexing is an attractive method due to polarization sensitivity of artificially tailored meta-atoms that can completely alter the polarization state of the interacting light. Polarization multiplexed metasurface holograms that support only two orthogonal states and based on spatial multiplexing with more than one meta-atom within a unit-cell have been described previously[30,31]. Such spatial multiplexing usually results in strong cross-talk between the information channels and lowers the space-bandwidth product. Furthermore, the increased space requirement for the larger unit cell reduces the information density, which might counterbalance the additional information capacity of the second polarization channel. By using birefringent metasurfaces composed of a single meta-atom per unit-cell that possess a variation in the phase shifts for the orthogonal polarization states, an alternative approach for realizing polarization multiplexing has been demonstrated[32,33].

Nevertheless, the full capacity of all polarization channels (input/output polarizations) has not yet been explored. This limits the improvement of the information capacity stored in metasurface holograms and restricts the versatile functionality of holographic-based optical devices.

Here, we demonstrate a novel method for realizing multichannel vectorial holography and show its potential for obtaining dynamic displays and high-security applications. We explore birefringent metasurfaces for the complete control of polarization channels with the freedom of designing both the polarization-dependent phase shift and polarization rotation matrix. We show that although the target holographic phase profiles have quantified phase relations, they can process very different information within different polarization manipulation channels. The reconstructed vectorial images can be switched with negligible cross-talk by selecting the desired combination of input/output polarization states. A set of reconstructed images can conceive different meanings and therefore host unique merits for encryption. Our demonstrated multiplexing method may lead to a new frontier for applications related to dynamic holographic displays, switchable optical devices, data storage, and optical encryption/anti-counterfeiting, which are compatible with most optical systems that operate in transmission. Furthermore, our proposed method can also be adapted to arbitrary spin to orbital angular momentum conversion or other types of polarization/phase modulation and enhance the information capacity.

**Results:**

We comparatively study two schemes of multiple polarization channels (from orthogonal two channels to twelve channels) by using birefringent dielectric metasurfaces, as shown in Figure 1. We show that polarization and angle-multiplexed holograms can be realized by utilizing nanofins with different cross-sections but without rotation (Fig. 1a). Two sets of off-axis holographic images can be generated with two orthogonal states. However, by considering the flexibility for designing the entire Jones matrix, we can achieve more multiplexed functionalities with complete control of polarization and phase (Fig. 1b). Three independent images (we chose the words "holography", "meta", and "surface" as the reconstructed images) can be reconstructed successfully with high resolution and high fidelity for different combinations of input and output polarization channels. All combinations of these three images in a total of twelve different polarization channels can be observed ("meta" + "holography", "meta" + "surface", "surface" + "holography" and "meta" + "surface" + "holography") with satisfactory efficiencies. In addition, the reconstructed images exhibit a vectorial feature for the polarization state of the different words in analogy to optical vector beams.

For achieving multichannel vectorial holography, we have to explore suitable building blocks for the metasurfaces. As is well known, the general relation between the electric field of the input ($E^{in}$) and output ($E^{out}$) waves at each pixel can be expressed using the Jones matrix description:

$$E^{out} = TE^{in} \quad \text{where} \quad T = \begin{bmatrix} T_{xx} & T_{xy} \\ T_{yx} & T_{yy} \end{bmatrix} \quad (1)$$

Indeed, any desired symmetric and unitary Jones matrix can be realized using a birefringent metasurface if the polarization-dependent phase shift ($\phi_x$, $\phi_y$) and the orientation angle $\theta$ can be chosen freely [33]:

$$T = V \begin{bmatrix} e^{i\phi_x} & 0 \\ 0 & e^{i\phi_y} \end{bmatrix} V^T = R(\theta)\Delta R(-\theta) \quad (2)$$

where $\Delta$ indicates the eigenvalue matrix of the Jones matrix $T$ and the real unitary matrix $V$ can be treated as a rotation matrix $R$. Crucially, the desired phases (combining both dynamic and geometric phases) can be extended to impart on any set of orthogonal polarization states.

We first consider the two-channel polarization multiplexed holograms as the simplest case. By formulating the Jones matrix of each pixel as:

$$T = \begin{bmatrix} e^{i\varphi_1} & 0 \\ 0 & e^{i\varphi_2} \end{bmatrix} \quad (3)$$

We can easily tailor the desired output light to $E^{out}_{x,inc} = \begin{bmatrix} e^{i\varphi_1} \\ 0 \end{bmatrix}$ in each pixel according to the profile of the hologram 1 while guaranteeing the phase distribution of

$E_{y,inc}^{out} = \begin{bmatrix} 0 \\ e^{i\varphi_2} \end{bmatrix}$. Therefore, the original image 1 and image 2 for orthogonal incident polarizations can be reconstructed and switched. These two channels for polarization multiplexed metasurface holograms can be easily realized by generating a polarization-dependent phase shift in the $x$ and $y$ directions but with an identical orientation angle for all nanofins. Note that to further enlarge the information capacity, one can also encode angular or distance multiplexed holographic images for each phase profile of $\varphi_1$ and $\varphi_2$.

More complex multiplexed functionalities can be realized by utilizing the design freedom of the rotation matrix $R$. Assuming that the Jones vector of the output light is described by $E_{x,inc}^{out} = \frac{\sqrt{2}}{2}\begin{bmatrix} e^{i\varphi_1} \\ e^{i\varphi_2} \end{bmatrix}$ when the incident light is $x$-polarized light. Now, the $x$ and $y$ components of the output light can be designed to reconstruct the original holographic image 1 and image 2 with equal intensity and high contrast, respectively. More importantly, the reconstructed images can be switched by selecting the desired transmission polarization with negligible polarization cross-talk. For such a situation, the Jones matrix of each pixel can be derived as:

$$T = \frac{\sqrt{2}}{2}\begin{bmatrix} e^{i\varphi_1} & e^{i\varphi_2} \\ e^{i\varphi_2} & e^{i(2\varphi_2-\varphi_1+\pi)} \end{bmatrix} \quad (4)$$

From the above Jones matrix, one can recognize that when the incident light polarization is changed to $y$-polarized light, the Jones vector of the output light will be expressed by $E_{y,inc}^{out} = \frac{\sqrt{2}}{2}\begin{bmatrix} e^{i\varphi_2} \\ e^{i(2\varphi_2-\varphi_1+\pi)} \end{bmatrix}$. This additional phase relation between $\varphi_1$ and $\varphi_2$ for the $y$-component of the output field opens up the possibility of encoding an additional holographic image. By utilizing a suitable algorithm for the generation of holograms, holographic image 3 can be reconstructed by satisfying the relation $\varphi_3 = 2 \times \varphi_2 - \varphi_1 + \pi$. Hence, the $y$-component of the output light results in reconstruction of image 3 by illumination with $y$-polarized light. Meanwhile, the concept can be extended further by utilizing circular polarization states for the incident wave. For circularly polarized input light, the output field can be derived as:

$$E_{L(R),inc}^{L(R),out} = \frac{e^{i\varphi_1} + e^{i(2\varphi_2-\varphi_1+\pi)}}{4}, \quad E_{L(R),inc}^{R(L),out} = \frac{e^{i\varphi_1} + 2e^{i(\varphi_2\pm\pi/2)} + e^{i(2\varphi_2-\varphi_1)}}{4} \quad (5)$$

where the super/subscript in the brackets denote the alternative case for orthogonal left-handedness (*L*) or right-handedness (*R*) circular polarization. Based on the phase relation of the output fields, one can find that three different independent images (1-3) and all combinations of these images (1+2, 1+3, 2+3, 1+2+3) can be reconstructed from each item using the above equations, while a constant phase difference of ±π/2 and 0/π in Eq. 5 will only result in the same image. Hence, we obtain twelve channels in total (different combinations of incident/output polarization: $T_{xx}$, $T_{xy}$, $T_{yx}$, $T_{yy}$, $T_{ll}$, $T_{lr}$, $T_{rl}$, $T_{rr}$, $E_x^{out}$, $E_y^{out}$, $E_L^{out}$, $E_R^{out}$) and seven combinations of holographic images within one single metasurface, without any additional spatial multiplexing.

To acquire the target phase profiles to fulfill Eq. 4 (we can view Eq. 3 as a special case without rotation), both the hologram generation algorithm and suitable design of the meta-atoms are key points. We build a new hologram algorithm based on a modified Gerchberg-Saxton (GS) scheme by considering the quantified relations of each other ($\varphi_1$, $\varphi_2$, and $\varphi_3=2\varphi_2-\varphi_1+\pi$) to adapt to each polarization channel. As a phase retrieval method, the desired phase distribution obtained with the GS algorithm is calculated by constructing an iterative loop between the object plane and the hologram via a Fourier transform[37]. Crucially, the phase $\varphi_3$ has a quantified relation with both other phases $\varphi_1$ and $\varphi_2$, while the latter two phase profiles can be altered independently. Therefore, we used another serial loop by connecting all of these three phase profiles. A flowchart with details can be found in the Supplementary Material. Meanwhile, we use the birefringence as well as the rotational property (adapted from Eq. 2) of the simplest rectangular nanofins with a tailorable cross-section and azimuthal angles. Such structures provide a suitable platform for the realization of multichannel vectorial holograms. To realize the desired phase shifts in each polarization channel, one has to guarantee that the phase shift of either $\varphi_1$ or $\varphi_2$ can cover the entire 0~2π range while still allowing all possible combinations of ($\varphi_1$, $\varphi_2$). Furthermore, we have to maintain a uniform amplitude distribution with high efficiency. A 2D parameter optimization by using a rigorous coupled wave analysis (RCWA) method was carried out. The simulation results for the amplitude and phase distribution of Si nanofin for the orthogonal polarization state illumination are shown in Figure 2 (see Methods).

In the following, we fabricated several designed dielectric silicon metasurfaces on top of a glass substrate by using a plasma etching process, followed by electron beam lithography for patterning (see Methods). The experimental setup and two typical scanning electron microscopy images of the samples with and without rotation are shown in Figure 3. For the specific simple case of no rotation (Eq. 3), we encode 4 separate original images into $\varphi_1$ and $\varphi_2$ by considering polarization and angular multiplexing, respectively, as shown in Figure 4. This metasurface yields off-axis holographic images of a cartoon tiger and a snowman with high fidelity and high resolution when illuminated by *x*-polarized light. By switching to *y* polarization for the incident light, the reconstructed images are changed to a teapot and a cup. Note that in this case, there are only two polarization channels, with both pairs of the holographic image reconstructed and made to disappear simultaneously by rotating the polarizer behind the sample. For the polarization of the incident light with an angle of 45° or 135°, all four original images can be observed simultaneously. The experimental results are in good agreement with the simulation results, which confirms the basic design principle. In our experiment, we define two types of diffraction efficiencies. The diffraction efficiency of each polarization channel ($T_{xx}$, $T_{xy}$, $T_{yx}$, and $T_{yy}$) is defined as the ratio of the power of the output light in the different channels to the power of the incident light. These diffraction efficiencies of our fabricated samples are retrieved from spectral analysis (transmission spectra). The corresponding experimental results are shown in the Supplementary Material. The net diffraction efficiency for the holography is defined as the ratio of the intensity of the single reconstructed image to the power of the incident light (the residual power in the central area is not included). For the four reconstructed images, the net diffraction efficiencies (cartoon tiger, snowman, teapot, cup) are 15.37%, 10.87%, 13.08%, 10.92%, respectively (see the video and further efficiency analysis in the Supplementary Materials).

Furthermore, we fabricate another dielectric metasurface hologram that can utilize additional polarization channels by considering the additional design freedom afforded by the orientation angles of the nanofins. In such a way, more complex multiplexing functionalities can be obtained. From the linear polarization experimental results (top two rows), we observe that when the incident light is *x*-polarized, the *x* ($E_{x,inc}^{x,out}$) and *y* ($E_{x,inc}^{y,out}$) components of the output field will contribute to the reconstruction of the words "holography" and "meta", respectively. Hence, the total electromagnetic field of the output light ($E_{x,inc}^{norm,out}$) results in reconstruction of both words ("holography" and "meta") simultaneously, but with orthogonal 'vectorial' properties for each word (Figure 5*g-i*). Upon switching to the incident light with *y* polarization, the *x* ($E_{y,inc}^{x,out}$) and *y* ($E_{y,inc}^{y,out}$) components of the output light will be switched to reconstruct the words "meta" and "surface", respectively (Fig. 5*j-l*). We further carried out verification of the multiplexed holograms under illumination with circularly polarized light (bottom two rows). When the incident light is LCP / RCP, we can observe the words "holography" and "surface" simultaneously by selecting the LCP ($E_{L,inc}^{L,out}$) / RCP ($E_{R,inc}^{R,out}$) components of the output light (Fig. 5*w*, 5*s*). All three images can be obtained by using the RCP ($E_{L,inc}^{R,out}$) / LCP ($E_{R,inc}^{L,out}$) components of the output light (Fig. 5*v*, 5*t*). Similarly, we can determine that the vectorial nature of the reconstructed images of the words "holography" and "surface" is linearly polarized from Eq. 5. While for the word "meta", it only appears in the orthogonal circular polarization channels (Fig. 5*x*, 5*u*), as can be derived from Eq. 5. Therefore, the polarization state of the reconstructed word "meta" is right (left) handedness circularly polarized. Further analysis of arbitrary elliptical input/output polarization combinations can be carried out that will result in a much more complex "vectorial" nature for the reconstructed holographic images. We determined the net diffraction efficiencies for the reconstructed images ("holography", "meta", and "surface") to be 15.97%, 8.03%, and 9.91% for the case of a linearly polarized combination, respectively. Therefore, three independent images and all the combinations of these images (twelve channels in total with seven different combinations) can be reconstructed with high contrast by our proposed method. Additional results for demonstrating the feasibility of achieving a dynamic holographic display and efficiency analysis can be found in the Supplementary Material.

In addition, the method also enables encrypting different original images that can be superimposed at the same spatial location. Such superposition has the ability to convey a different meaning in the reconstructed image and can be used to provide alterable information content for the image, i.e., encryption, as shown in Figure 6. Here, we chose a dice as our original image. The dice has six faces that are represented by the number of pips on the surface. Interestingly, by using a suitable selection of different combinations of input/output polarization states, a different number of pips (from one to six) can be observed. This kind of illusion for viewing different sides of the dice results from the increased multiplexing capability of our method that can encode up to six different images for the various combinations of polarization states (see the video in the Supplementary Materials).

**Discussion:**

Our demonstrated multiplexing algorithm can result in a dynamic vectorial holographic display and encryption. Only by using the correct polarization keys can the receiver obtain the exact information delivered (as for the dice case). An even higher flexibility can be obtained by further increasing the complexity of the images (for example, by using more than two dice as the original images or using angle/distance/hybrid multiplexing in each independent phase profile) together with a detailed analysis of the reconstructed vectorial image properties. Such a metasurface with a compact physical size is easy to hide and transport. Even if the metasurface eavesdrops, the correct polarization combination (input/output) together with the additional vectorial nature of each image can provide a triple protection security lock for the communication information, which will greatly enhance the complexity of the decryption.

Such multichannel vectorial holography also exhibits a relatively large working bandwidth. This phenomenon can be understood in the following manner: due to the robustness of our silicon metasurface holograms and the fact that holograms generated by the GS algorithm are wavelength independent, together with the small variation of the Si refractive index, the reconstructed images can be observed with high fidelity even away from our design wavelength of 800 nm by using dielectric metasurfaces. In our experiment, we can achieve broadband reconstruction from the near-infrared region to visible light (between 600 nm and 800 nm) with or without rotation of the nanofins. The broadband response arises from the robustness of our silicon metasurface holograms and the fact that the holograms generated by the GS algorithm are wavelength independent. Detailed analysis through consideration of the phase shift caused by different wavelengths and statistics number for each nanofin over the entire metasurfaces together with reconstruction of the Jones matrix to accurately predict the broadband effect from simulation and experimental verification can be found in the Supplementary Material. Such a phenomenon can demonstrate the robustness of the metasurface holograms and can be further applied to realize color images. Alternatively, Titanium Dioxide ($TiO_2$) or Silicon Nitride (SiN) can be used to realize such holograms in the visible range with larger efficiency.

Information and image multiplexing based on the polarization states of light is a powerful tool that can be easily realized with optical metasurfaces. Here, we have demonstrated a novel method for achieving multichannel vectorial holography based on birefringent all-dielectric metasurfaces to explore the full capacity of polarization. By considering the birefringent anisotropy property and extra design freedom afforded by the rotation matrix, together with the use of smart multiplexing algorithms for establishing quantified related phase profiles, high-dimensional multichannel polarization multiplexed holograms were successfully realized using a simple nanofin as the building block. Compared to previously demonstrated metasurface multiplexed holograms that can only be switched between two states by incident light with two orthogonal polarization states, here, all combinations within twelve polarization channels in total can be obtained. Further application of arbitrary spin to angular momentum conversion and various phase modulation/beam shaping can be realized accordingly.

**Materials and methods:**

**Fabrication of the metasurfaces**

The dielectric metasurfaces were fabricated on a glass substrate following the processes of deposition, patterning, lift off and etching. First, a 600-nm-thick amorphous silicon (a-Si) film was deposited by plasma enhanced chemical vapor deposition (PECVD). A poly-methyl-methacrylate resist layer was spin-coated onto the a-Si film and baked on a hot plate at 180 °C for 2 min to remove the solvent. Then, the desired structure was patterned by using standard electron beam lithography and subsequent development in 1:3 MIBK:IPA solution. Next, the sample was washed with IPA before being coated with a 45-nm-thick chromium layer by electron beam evaporation. Next, a lift-off process in hot acetone was performed. Finally, by using inductively coupled plasma reactive ion etching (ICP-RIE), the desired structure was transferred from chromium to silicon. The metasurfaces are composed of 1000×1000 nanofins (lattice constant along the *x*- and *y*-axis of 400 nm) with different cross-sections (width and length in the range of 80 ~ 280 nm, height 600 nm) and orientation angles. To minimize fabrication errors and to obtain high-quality nanostructures, we took several precautions. Before fabricating the actual metasurface with all the various types of nanopillars, we first optimized the nanopillars with the smallest size (with and without rotation) for the electron beam lithography and the plasma etching process. After the best parameters were obtained for each nanopillar size, we used these values for the fabrication of the actual metasurface sample. We found that the structure size deviation is below five percent compared to the target values.

**Design and numerical simulations**

Our birefringent dielectric metasurfaces are designed using amorphous silicon nanofins on top of a glass substrate, as shown in Fig. 2a. To achieve the desired phase shifts ($\varphi_x$, $\varphi_y$), we carry out a 2D parameter optimization by using a rigorous coupled wave analysis (RCWA) method. The length $L$ and width $W$ of the nanofin are both swept in the range of 80 to 280 nm, while maintaining the height $H$ at 600 nm and the period size $P$ at 400 nm. The values for $H$ and $S$ are carefully chosen to guarantee that the phase of the output light can cover $0\sim2\pi$ and eliminate undesired orders of diffraction. For the simulation, the nanofin is placed onto a glass substrate ($n_{SiO2}$=1.5). The wavelength of incident light is fixed at 800 nm, and the corresponding refractive index of amorphous silicon is $n_{Si}$=3.6941+0.0065435*1i. The obtained transmission amplitudes and phases for the two transmission coefficients $t_{xx}$ and $t_{yy}$ are shown in Fig. 2b-d. We found that the transmission amplitudes for most of the nanofins with different cross-sections are over 90%. Note that the orientation angles of the nanofins are determined by Eq. 2 and Eq. 4 for the multichannel polarization multiplexing.

**Optical measurement**

For the optical characterization of the performance of the metasurface holograms, we use the setup that is shown in Fig. 3. A pair of linear polarizers or a combination of linear polarizers and quarter-wave plates is placed in front of and behind the sample to select the desired incident/transmitted linearly/circularly polarized beam. Our metasurface sample is placed in the focal plane of an objective lens (40×/NA=0.6) to guarantee that the Fourier plane is located in the back focal plane. Meanwhile, the magnifying ratio and numerical aperture of the objective lens are carefully chosen for the purpose of collecting all the diffraction light from the sample and reconstructing the holographic images in the Fourier plane. Another objective/lens is used for capturing the Fourier plane on a CCD camera.

**Acknowledgments**

The authors acknowledge the funding provided by the National Key R&D Program of China (No. 2017YFB1002900) and the European Research Council (ERC) under the European Union's Horizon 2020 research and innovation programme (grant agreement No. 724306). L.H. acknowledge the support from the National Natural Science Foundation of China (No. 61775019) program, the Beijing Municipal Natural Science Foundation (No. 4172057), the Beijing Nova Program (No. Z171100001117047) and the Young Elite Scientists Sponsorship Program by CAST (No. 2016QNRC001).

**Author contributions**

L.H. and T.Z. proposed the idea, L.H. and R.Z. conducted pattern designs and numerical simulations, Q.W. and R.Z. conducted the hologram generation, B.S. and C.T. fabricated the samples, T.Z., L.H., R.Z. and B.S. performed the measurements, L.H., T.Z., R.Z., and B.S. prepared the manuscript. L.H., Y.W., and T.Z. supervised the overall project. All of the authors analyzed the data and discussed the results.

**Conflict of interests**

The authors declare that they have no conflicts of interest.

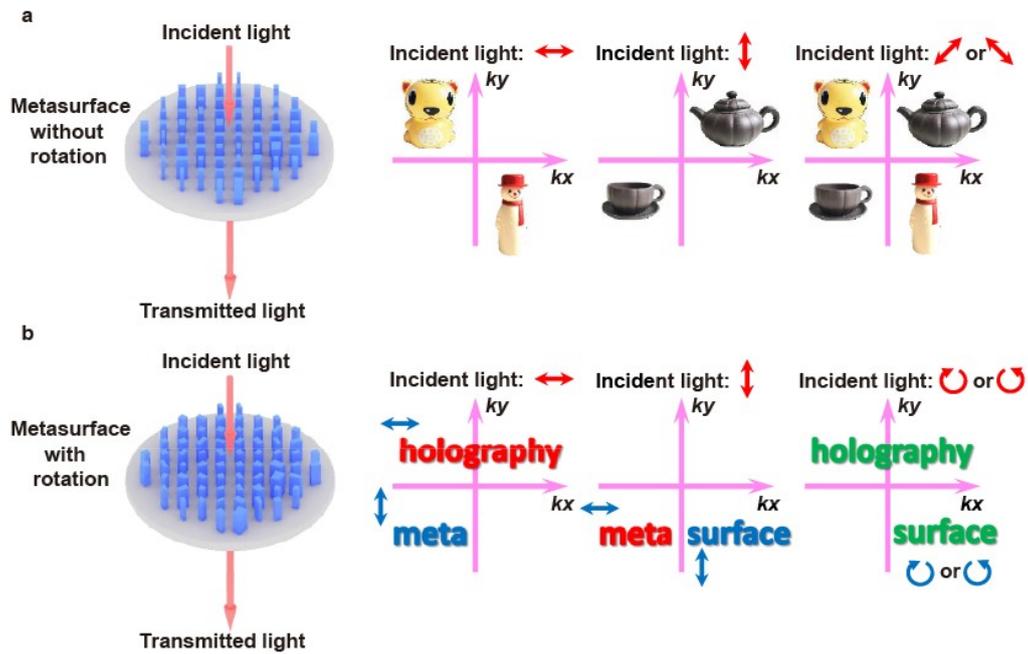

**Fig. 1 Principle of our designed metasurface hologram.** Schematic illustrations of polarization multiplexed holograms based on dielectric metasurfaces. The red and blue arrows indicate the polarization of the incident light and the transmission axis of the polarizer placed behind the metasurface sample. The red, blue and green color of the reconstructed images (the words "holography", "meta" and "surface") represent the *x*, *y* and RCP (or LCP, with the same helicity as incident light) component of the output light, respectively. **a** Two channel polarization and an angle multiplexed hologram based on metasurfaces composed of nanofins with different cross-sections but fixed orientation angles, which can be used to reconstruct two sets of off-axis images. **b** Multichannel polarization multiplexed hologram based on metasurfaces composed of nanofins with different cross-sections and orientation angles, which can be used to reconstruct three independent images and all combinations of these images (twelve channels in total).

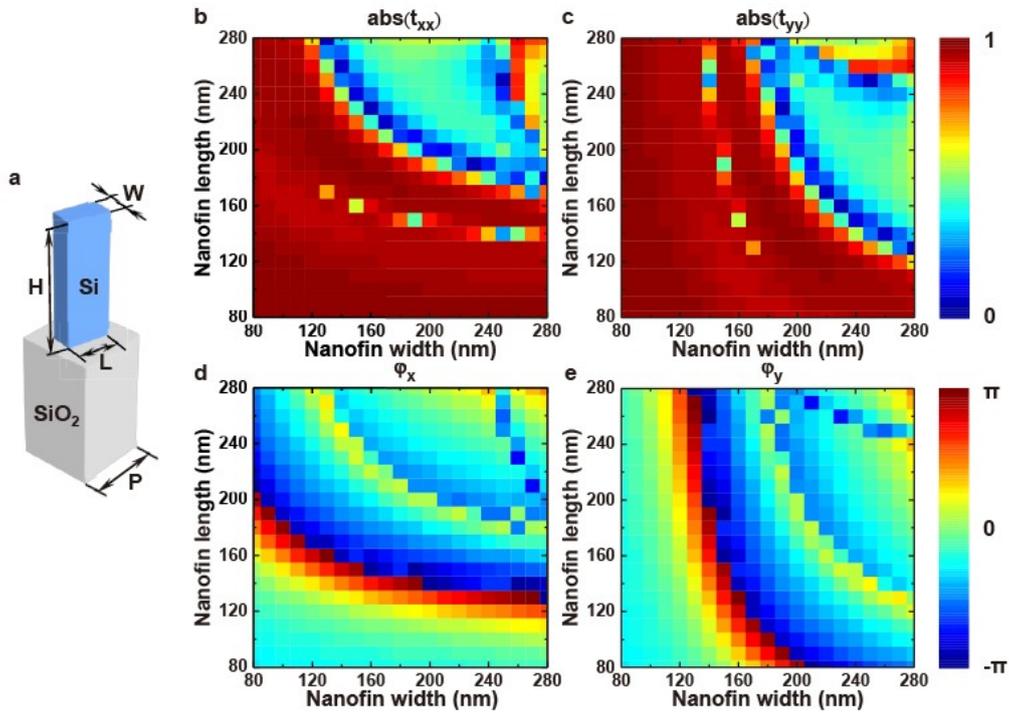

**Fig. 2 Simulated amplitude and phase of the transmission coefficients of our designed silicon nanofin**. **a** Schematic illustration of an amorphous silicon nanofin positioned on a glass substrate. The metasurface will be composed of a periodic arrangement of such unit-cells. **b-e** Simulation results for the amplitude and phase of the transmission coefficients $t_{xx}$ and $t_{yy}$ shown for a 2D parameter optimization by using a rigorous coupled wave analysis method. The length and width of the nanofin are both swept in the range of 80 to 280 nm at an incident wavelength of 800 nm.

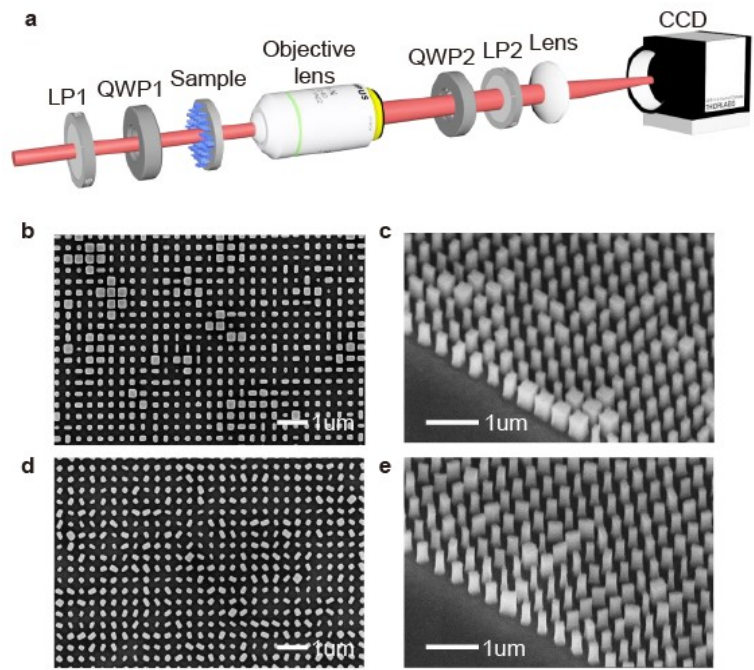

**Fig. 3 Experimental setup and scanning electron microscopy images of the fabricated metasurface samples. a** The experimental setup for the observation of the holographic images. The two linear polarizers (LP1, LP2) and two quarter-wave plates (QWP1, QWP2) are used to set the precise polarization combination for the incident/transmitted light. The lens images the back focal plane of the microscope objective lens (40x/0.6) to a CCD camera. **b-e** Scanning electron microscopy images of two typical fabricated silicon metasurface samples shown with a top and side view. The metasurface holograms are composed of 1000×1000 nanofins with different cross-sections and orientation angles.

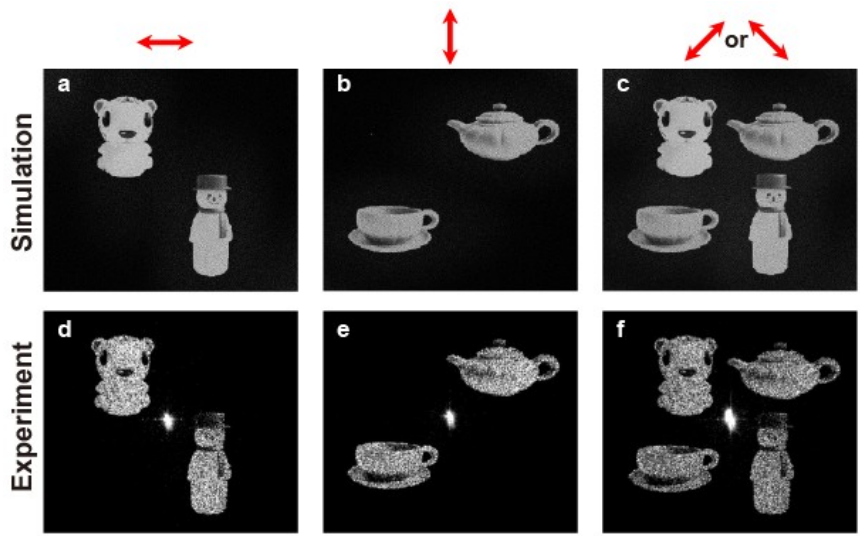

**Fig. 4 Simulated and experimental results for the two channel polarization and angle multiplexed hologram. a-f** Two sets of off-axis images are reconstructed, which can be switched by changing the polarization of the incident light (denoted by the red arrows).

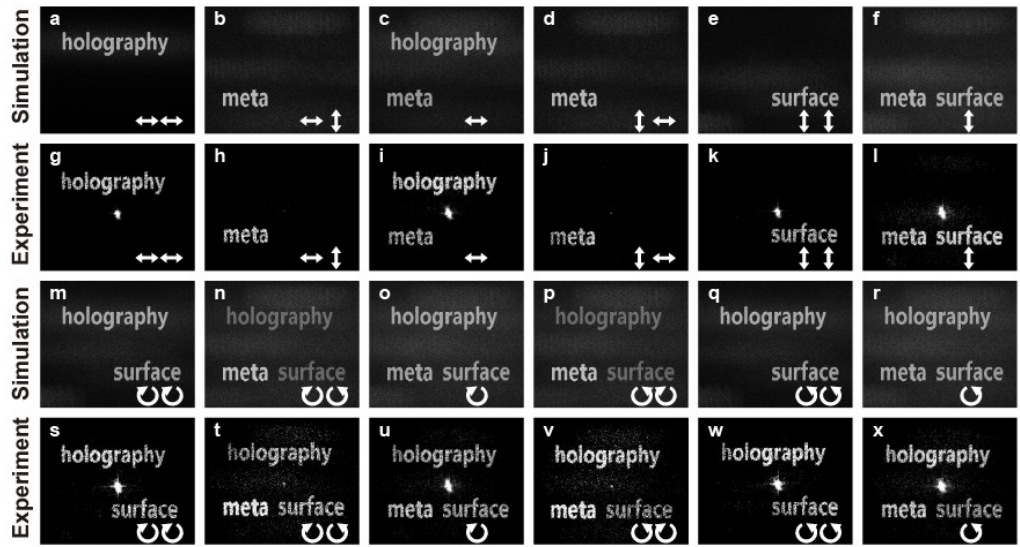

**Fig. 5 Simulated and experimental results for the multichannel vectorial holography (linear and circular channels). a-x** The two white arrows in the corners indicate the input (first arrow) and output (second arrow) polarization of light. In total, twelve polarization channels and seven different image combinations are demonstrated. The vectorial nature of each word can be analyzed from Eq. 4 and Eq. 5.

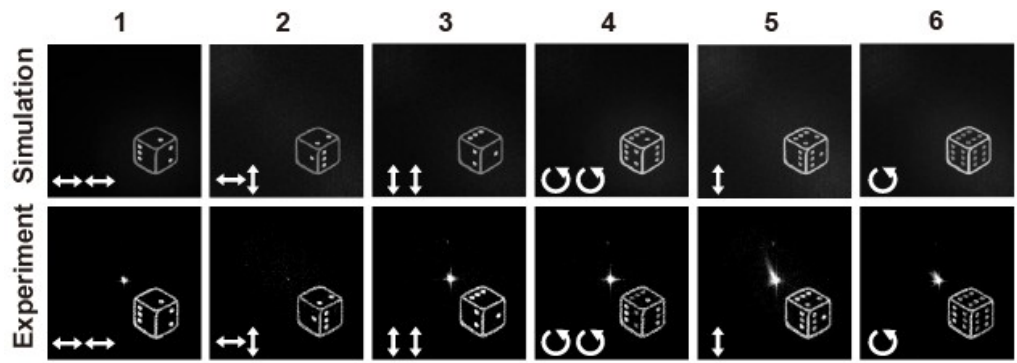

**Fig. 6 Simulated and experimental results for the reconstruction of a dice with different exposed faces.** The points one to six are reconstructed by using different combinations of input/output polarization. The two white arrows in the corners indicate the input (first arrow) and output (second arrow) light polarization. For numbers 5 and 6, no polarization analyzer was used.